\pdfoutput=1 
\documentclass{JINST}

\title{Charge Collection Efficiency Simulations of Irradiated Silicon Strip Detectors}

\author{Timo Peltola\thanks{Speaker.}~\thanks{On behalf of the CMS Tracker collaboration.}~ \\
\llap{}Helsinki Institute of Physics\\
P.O.Box 64 FI-00014 University of Helsinki, Finland\\
E-mail: \email{timo.peltola@helsinki.fi}}

\abstract{During the scheduled high luminosity upgrade of LHC, the world's largest particle physics accelerator at CERN, the position sensitive silicon detectors installed in the vertex and tracking part of the CMS experiment will face a more intense radiation environment than the present system was designed for. Thus, to upgrade the tracker to the required performance level, comprehensive measurements and simulation studies have already been carried out. 

Essential information of the performance of an irradiated silicon detector is obtained by monitoring its charge collection efficiency (CCE). From the evolution of CCE with fluence, it is possible to directly observe the effect of the radiation induced defects on the ability of the
detector to collect charge carriers generated by traversing minimum ionizing particles (MIPs).

In this paper the numerically simulated CCE and CCE loss between the strips of irradiated silicon strip detectors are presented. The simulations based on the Synopsys Sentaurus TCAD framework were performed before and after irradiation for fluences up to $1.5\times10^{15}$ $\textrm{n}_{\textrm{\tiny eq}}$cm$^{-2}$ for n-on-p sensors. A two level defect model and non-uniform three level defect model were applied for the proton irradiation simulations, and a two level
model for neutrons. The results are presented together with the measurements of strip detectors
irradiated by different particles and fluences and show considerable agreement for both CCE and its position dependency. 
}

\keywords{Si microstrip and pad detectors; Detector modelling and simulations II; Radiation damage to detector materials (solid state)}

\begin{document}

\section{Introduction}\label{sec:1}
\paragraph{} Position sensitive silicon detectors are largely employed in High Energy Physics (HEP) experiments due to their outstanding performance. They are currently installed in the vertex and tracking part of the CMS experiment at LHC, the world's largest particle physics accelerator at CERN, Geneva.

The foreseen upgrade of the LHC accelerator at CERN, namely the high luminosity (HL) phase of the LHC (HL-LHC scheduled for 2023), will enable the use of maximal physics potential of
the facility. After 10 years of operation, the integrated luminosity of 3000 fb$^{-1}$, resulting in expected fluence above 10$^{15}$ $\textrm{n}_{\textrm{\tiny eq}}$cm$^{-2}$ for strip sensors $\sim$20 cm from vertex, will expose the tracking system at HL-LHC to a radiation environment that is beyond the capability of the present system design. Thus, an upgrade of the all-silicon central trackers is required. This will include higher granularity and radiation hardness for the sensors to be able to withstand higher radiation levels and higher occupancies also in the innermost layers closest to the interaction point.
For the upgrade, extensive measurements and simulation studies for silicon sensors of different designs and materials with sufficient radiation tolerance have been initiated.

Complementing measurements, simulations are an integral part of the R$\&$D of novel silicon radiation detector designs with upgraded radiation hardness. When numerical simulations are able to verify experimental results, they will also gain predictive power. This can lead to reduced time and cost budget in detector design and testing. The charge collection efficiency (CCE) can be considered as one of the essential properties for the determining of the radiation hardness of a silicon detector due to the direct information gained on the effect of radiation induced defects on the ability of a detector to collect charge carriers generated by traversing particles. Hence, its accurate reproduction by simulation can be a valuable tool in the development of radiation hard strip sensors. 

The simulations presented in this paper are a continuation of the study reported in \cite{bib1} and likewise in that paper were carried out using the Synopsys Sentaurus\footnote{http://www.synospys.com} finite-element Technology Computer-Aided Design (TCAD) software framework. For the charge collection efficiency simulations, separate two level defect models \cite{bib2} were applied for proton and neutron irradiated sensors. A non-uniform three level defect model \cite{bib1} was employed for the simulations of the position dependency of the CCE in proton irradiated sensors. The results are compared with measurements of Hamamatsu Photonics K. K. (HPK) produced strip detectors
irradiated with different particle species and fluences.

\section{Simulated radiation damage in silicon strip sensors}\label{sec:2}
\paragraph{} Radiation levels above $\sim10^{13}$ $\textrm{n}_{\textrm{\tiny eq}}$cm$^{-2}$ cause damage to the silicon crystal structure. Fluences above $1\times10^{14}$ $\textrm{n}_{\textrm{\tiny eq}}$cm$^{-2}$ lead to significant degradation of the detector performance. Defects are introduced both in the silicon substrate (bulk damage) and in the SiO$_2$ passivation layer, which affect the sensor performance through the interface with the silicon bulk (surface damage). Bulk damage degrades detector operation by introducing deep acceptor and donor type trap levels \cite{bib3}. The effects of the trapping includes higher leakage current, decreased CCE (and its position dependency) and a change in the effective doping concentration of the substrate. Surface damage consists of a positively charged layer accumulated inside the oxide and of interface traps that may be created close to the interface with silicon bulk \cite{bib4}. High oxide charge densities are detrimental to the detector performance since the electron layer generated under the SiO$_2$/Si interface can cause very high electric fields near the p$^+$ strips in p-on-n sensors and loss of position resolution in n-on-p sensors by providing a conduction channel between the strips.

In the simulation the bulk damage is approximated by an effective two-defect model that is tuned from the EVL-model \cite{bib5}. Separate capture cross sections and defect concentrations are used for proton and neutron irradiated sensors \cite{bib2} to match the leakage current $I_{\textrm{\tiny leak}}$, full depletion voltage $V_{\textrm{\tiny fd}}$ and transient current with experimentally obtained results. Surface damage is modelled in the simulation by placing a fixed charge $Q_{\textrm{\tiny f}}$ at the SiO$_2$/Si interface.

To study the position dependency of the CCE (CCE($x$)), a three level defect model close to the sensor surface \cite{bib2} is applied to reach the observed charge losses when the charge is injected between the strips \cite{bib6}, \cite{bib7}.

\section{Strip sensor simulation set-up}\label{sec:3}
\paragraph{} To compare simulations with measurements, the segmented sensor structures produced within the HPK campaign \cite{bib8} have been simulated following the real device features as close as possible.

\begin{figure}[tbp] 
\centering 
\includegraphics[width=.7\textwidth]{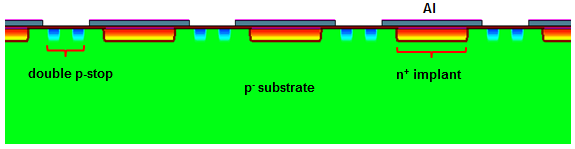}
\caption{Simulated n-on-p 5-strip sensor front surface with double p-stop strip isolation structure (not to scale). Colors: p$^+$/n$^+$ doping (blue/red), p$^-$ bulk doping (green), oxide (brown) and  Al (gray). The aluminum layer is repeated at the backplane where the reverse bias voltage is applied.}
\label{fig:1}
\end{figure}

For the simulations of n-on-p strip sensors with various strip width to pitch ratios and active thicknesses, a 5-strip structure was chosen to avoid any non-uniformities from border effects on the mesh formation at the center part of the device. Pictured in figure~\ref{fig:1} is a 5-strip n-on-p sensor with isolation between strips provided by a double p-stop implantation. The simulated strip sensor configurations were regions 5 (pitch = 120 $\mu\textrm{m}$, strip width = 41 $\mu\textrm{m}$, implant width = 28 $\mu\textrm{m}$) and 7 (pitch = 80 $\mu\textrm{m}$, strip width = 31 $\mu\textrm{m}$, implant width = 18 $\mu\textrm{m}$) from the MSSD structures used in the HPK campaign, with a strip length of 3.049 cm \cite{bib8}. For the n-on-p sensors the p-stop widths were 4 $\mu\textrm{m}$ and the spacing between p-stops was 6 $\mu\textrm{m}$, while the peak doping of $1\times10^{16}$ $\textrm{cm}^{-3}$ decayed to the bulk doping level within 1.5 $\mu\textrm{m}$, having equal depth with n$^+$ strip implants. 
The thickness of the oxide on the front surface was 250 nm and Al layers on the front and backplanes were 500 nm thick. Bulk dopings varied between $(3.0 - 3.4)\times10^{12}$ $\textrm{cm}^{-3}$ and the heavily doped boron and phosphorus implantations on the front and backplane, respectively, had the peak concentrations $5\times10^{18}$ $\textrm{cm}^{-3}$. The sensors had a physical thickness of 320 $\mu\textrm{m}$, but the active thicknesses were 300 $\mu\textrm{m}$ and 200 $\mu\textrm{m}$, as the active thickness was produced by a deep diffusion doping profile, as in real HPK produced sensors. Each strip had a biasing electrode at zero potential as well as AC-coupled charge collecting contacts. The reverse bias voltage was provided by the backplane contact.
%
\section{Charge collection efficiency simulations}\label{sec:4}
\paragraph{} For the CCE simulations, two-defect models for proton and neutron irradiations were applied. The results are compared with the CCE data measured by the ALiBaVa \cite{bib9} and the Silicon Beam Telescope (SiBT) \cite{bib10} set-ups.

\subsection{Method}\label{sec:4a}
\begin{figure}[tbp] 
\centering
\includegraphics[trim=0.1cm 0.1cm 0.175cm 0.15cm, clip=true, width=.6\textwidth]{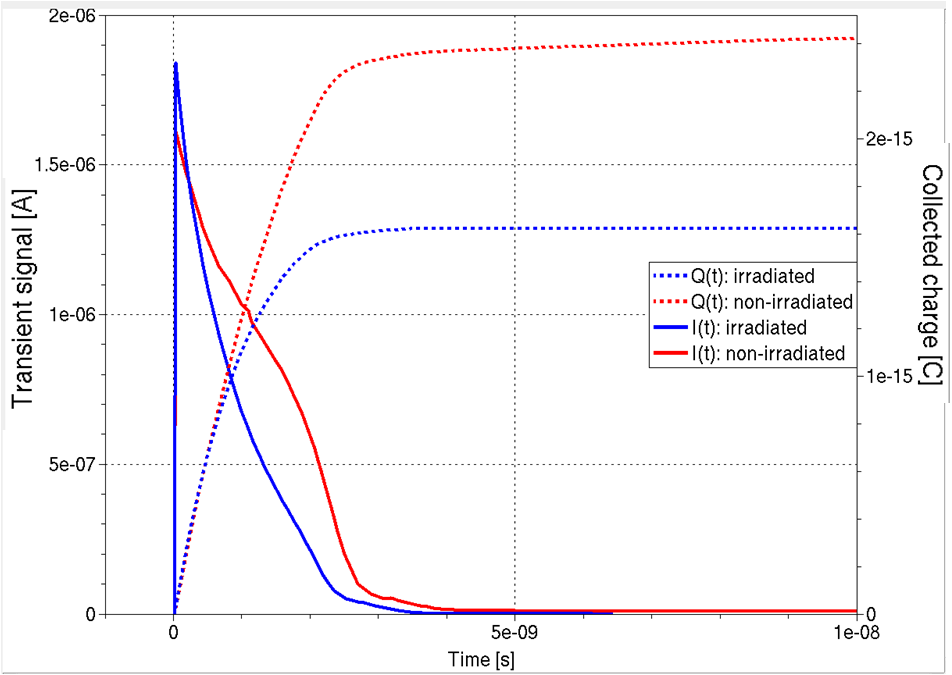}
\caption{The method to determine the CCE of an irradiated detector. The collected charge (right y-axis) is the integral of the transient signal (left y-axis) over time. The collected charge of an irradiated detector is a measure of efficiency relative to the non-irradiated detector.}
\label{fig:2}
\end{figure}

\paragraph{} Presented in figure~\ref{fig:2} are the transient currents and the collected charges of a 200 $\mu\textrm{m}$ active thickness strip sensor (described in section~\ref{sec:3}) before and after irradiation. Simulated CCE is defined as the ratio of the charge collected by an irradiated detector to the collected charge of a non-irradiated detector. As for the measurement of real detectors, the simulation temperatures were room temperature for the non-irradiated
and 253 K or 273 K for irradiated devices. Since both defect models are tuned to $T$ = 253 K, they were retuned for the higher temperature to maintain correct $I_{\textrm{\tiny leak}}$ by modifying the capture cross sections $\sigma_{\textrm{\tiny e,h}}$(273 K) = $0.75\times\sigma_{\textrm{\tiny e,h}}$(253 K). Charge injection was simulated by a MIP or infrared laser. The injection point for the CCE simulations was in the middle of the centermost strip and the mesh size around the trajectory was decreased accordingly to maintain convergence of the simulation. The collected charge was defined as the cluster charge from all five strips.

%

\subsection{Results and comparison with measurements}\label{sec:4b}
\begin{figure}[tbp] 
\centering
\includegraphics[width=.6\textwidth]{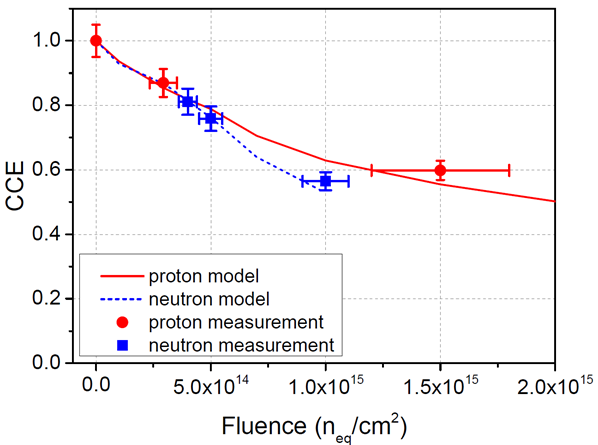}
\caption{Measured and simulated evolution of the CCE with fluence for 300 $\mu\textrm{m}$ active thickness n-on-p strip sensors for $V$ = -1 kV at $T$ = 253 K. The strip pitch is 80 $\mu\textrm{m}$ and implant width is 18 $\mu\textrm{m}$. Two interface charge values were used, $5\times10^{11}$ and $7\times10^{11}$ $\textrm{cm}^{-2}$ for fluences below and above $7\times10^{14}$ $\textrm{n}_{\textrm{\tiny eq}}\textrm{cm}^{-2}$, respectively. Experimental data is from the ALiBaVa set-up \cite{bib2}, \cite{bib12}, \cite{bib13}.}
\label{fig:3}
\end{figure}

\begin{figure}[tbp] 
\centering
\includegraphics[width=.48\textwidth]{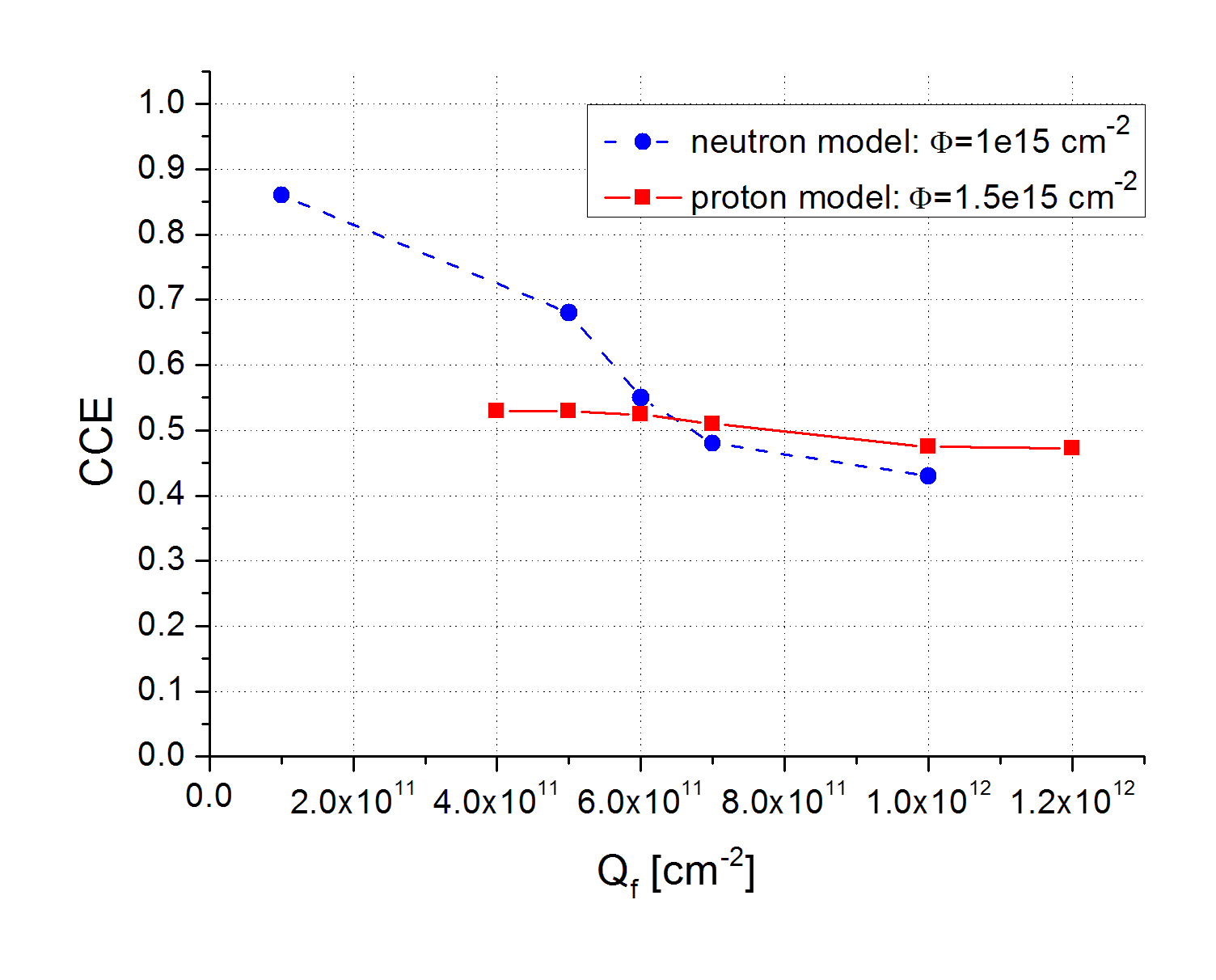}
\includegraphics[width=.48\textwidth]{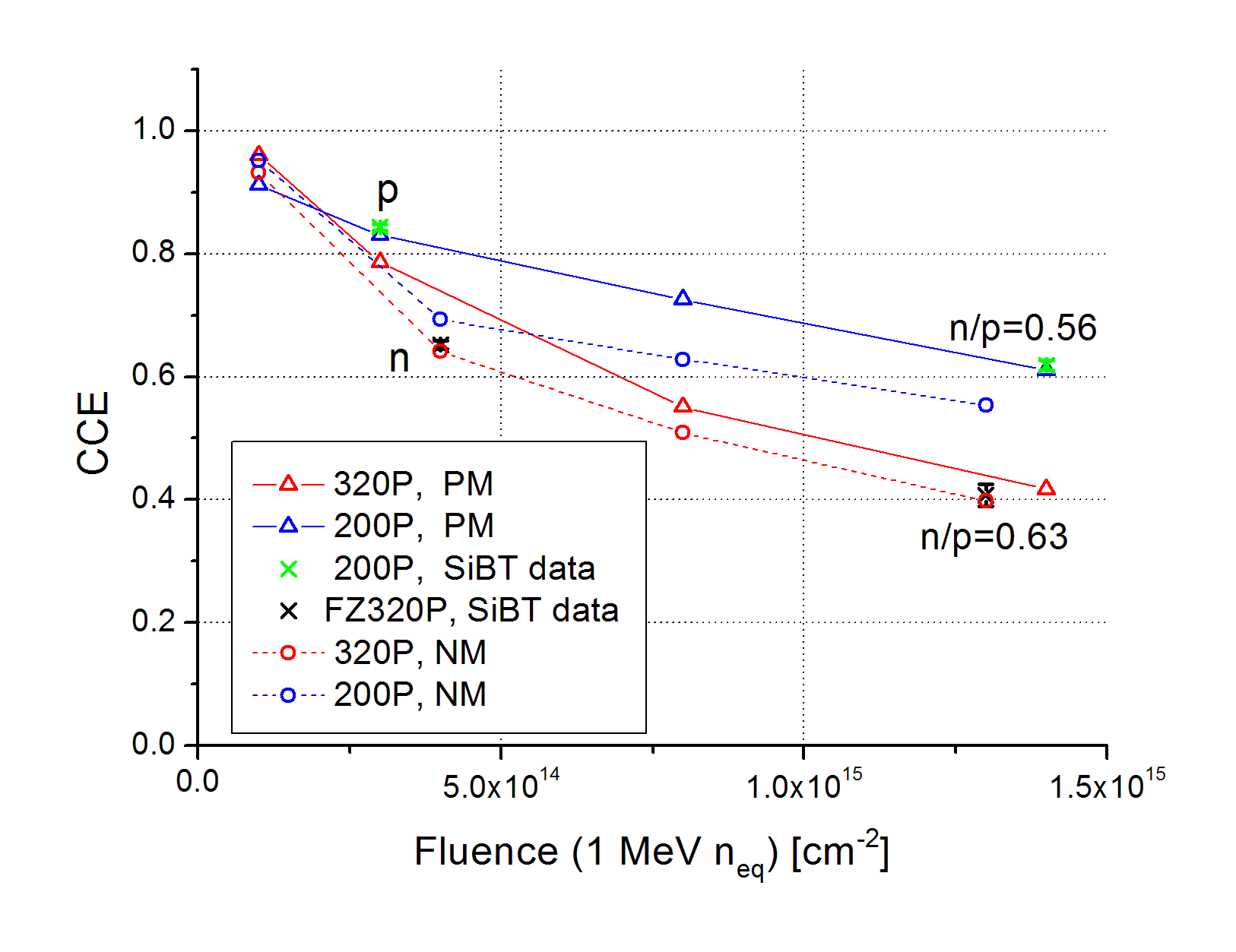}
\caption{(left) The dependency of the simulated CCE on the interface charge density $Q_{\textrm{\tiny f}}$ for 300 $\mu\textrm{m}$ active thickness n-on-p strip sensors at $V$ = -1 kV and $T$ = 253 K. The strip pitch is 80 $\mu\textrm{m}$ and implant width is 18 $\mu\textrm{m}$. (right) Measured and simulated CCE($\Phi_{\textrm{\tiny eq}}$) for n-on-p strip sensors at $V$ = -700 V ($V$ = -1 kV for the highest proton model dose in 200P sensor) and $T$ = 273 K. The strip pitch is 120 $\mu\textrm{m}$ and implant width is 28 $\mu\textrm{m}$. Irradiation types are marked in the plot as p = proton, n = neutron and n/p = mixed fluence with ratios of particles indicated. PM = proton model, NM = neutron model. Experimental data were measured with the SiBT set-up.}
\label{fig:4}
\end{figure}

\paragraph{} As shown in figure~\ref{fig:3} the simulated CCE of a 300 $\mu\textrm{m}$ active thickness sensor compares well to experimental data from both neutron and proton irradiated strip detectors.  
However, the maximum $Q_{\textrm{\tiny f}}$ values are kept considerably lower than expected in a real sensor for the highest proton irradiation fluence ($1 - 2\times10^{12}$ $\textrm{cm}^{-2}$ \cite{bib4}, \cite{bib11}) to ensure strip isolation in the proton model simulation. The dependency of the simulated CCE on $Q_{\textrm{\tiny f}}$ is further investigated in the left plot of figure~\ref{fig:4} for the highest neutron and proton fluences in figure~\ref{fig:3}. It is observed that the CCE changes in the studied $Q_{\textrm{\tiny f}}$ range for neutrons by about 40$\%$ and for protons within $\sim$10$\%$. Hence, if the $Q_{\textrm{\tiny f}}$ is set too low the charge multiplication due to high electric fields increase the CCE eventually to unphysically high values, especially for the neutron model, and if $Q_{\textrm{\tiny f}}$ is set too high, the undepleted region extending from the front surface results in too low CCE.

Reversely, to some extent the $Q_{\textrm{\tiny f}}$ can be used as a further tuning parameter of the CCE. This approach is applied on the right side of figure~\ref{fig:4} where $Q_{\textrm{\tiny f}}$ is iterated to find matching CCE with measurement for similar sensor structures and equal irradiation type. Two further CCE points are simulated with inter-/extrapolated $Q_{\textrm{\tiny f}}$ values using linear approximation. With these fixed $Q_{\textrm{\tiny f}}$ values, presented in table~\ref{tab:1}, a prediction of CCE($\Phi$) for sensors with different active thicknesses and equal irradiation type can then be made.

\begin{table}[tbp]
\caption{Interface charge densities used for the CCE simulations on the right side of figure~\protect\ref{fig:4}. PM = proton model, NM = neutron model.
}
\label{tab:1}
\smallskip
\centering
\begin{tabular}{lclclc}
    \hline
    {\small \bf{$\Phi_{\textrm{\tiny eq}}$ \textnormal{[cm$^{-2}$]}}} & {\small \bf{$Q_{\textrm{\tiny f}}$(NM) \textnormal{[cm$^{-2}$]}}} & {\small \bf{$Q_{\textrm{\tiny f}}$(PM) \textnormal{[cm$^{-2}$]}}}\\ 
    \hline
    {\small 1$\times$10$^{14}$} & {\small 6$\times$10$^{10}$} & {\small 1.4$\times$10$^{11}$}\\ 
    {\small 3$\times$10$^{14}$} & {\small -} & {\small 3$\times$10$^{11}$}\\
    {\small 4$\times$10$^{14}$} & {\small 9$\times$10$^{10}$} & {\small -}\\
    {\small 8$\times$10$^{14}$} & {\small 3.25$\times$10$^{11}$} & {\small 7.1$\times$10$^{11}$}\\ 
    {\small 1.3$\times$10$^{15}$} & {\small 6$\times$10$^{11}$} & {\small -}\\
    {\small 1.4$\times$10$^{15}$} & {\small -} & {\small 1.2$\times$10$^{12}$}\\
    \hline
\end{tabular}
\end{table}
%
\section{Simulated position dependency of CCE}\label{sec:5}
\paragraph{} The significant increase of surface damage with fluence in proton irradiated detectors requires the application of additional traps close to the surface to reach the experimentally observed strip isolation and CCE loss between the strips in n-on-p sensors. Hence, for the simulations of CCE($x$) to match with measurements, a non-uniform 3-level model needs to be applied. The results are compared with the CCE data measured with the SiBT setup.
\subsection{Method}\label{sec:5a}
\paragraph{} The experimentally observed CCE loss between the strips was simulated by first varying the position of the charge injection from the middle
of the pitch (maximum distance from the strip) to the center of the strip. Then the CCE loss was determined as the ratio of the difference in the collected cluster charge when the charge is injected at the center of the strip and in the middle of the pitch, to the collected cluster charge when injection is done at the center of the strip. As before, the cluster charge is determined as the sum of the charges collected on all five strips. A detailed description of the procedure can be found in \cite{bib2}. 
 
\begin{figure}[tbp] 
\centering
\includegraphics[width=.6\textwidth]{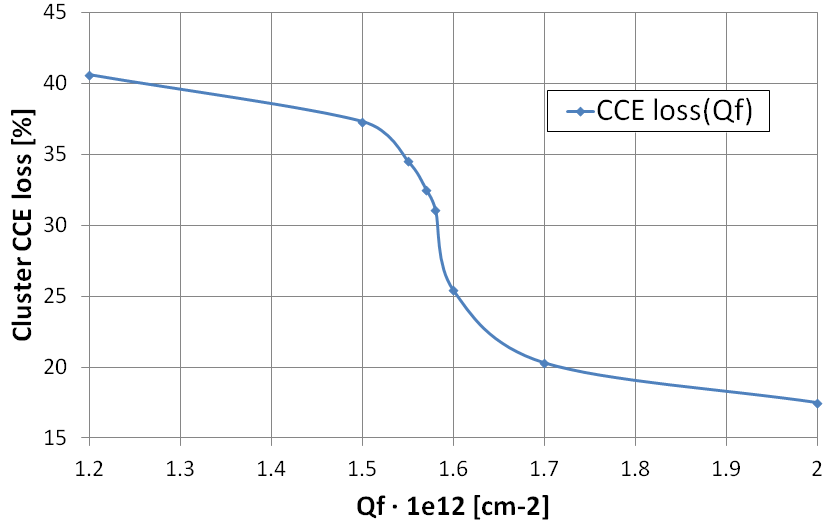}
\caption{Tuning of the simulated cluster CCE($x$) for a fixed concentration of shallow acceptors and voltage. Values of the interface charge density are scanned to reach a CCE loss value matching to measurement.}
\label{fig:5}
\end{figure}

The cluster CCE loss between the strips is then tuned to find agreement with the measured value by scanning the interface charge values, this process is presented in figure~\ref{fig:5}. The high CCE loss in the low $Q_{\textrm{\tiny f}}$ region in figure~\ref{fig:5} can be interpreted as the result of high trapping probability of the signal electrons due to large number of traps in the negative space charge region. On the other hand, the low CCE loss at high $Q_{\textrm{\tiny f}}$ region may be viewed as the result of the high number of accumulation layer electrons being able to fill the traps to the extent that decreases significantly the trapping probability of the signal electrons before they are collected at the strips.
\subsection{Results and comparison with measurements}\label{sec:5b}
\begin{figure}[tbp] 
\centering
\includegraphics[width=.487625\textwidth]{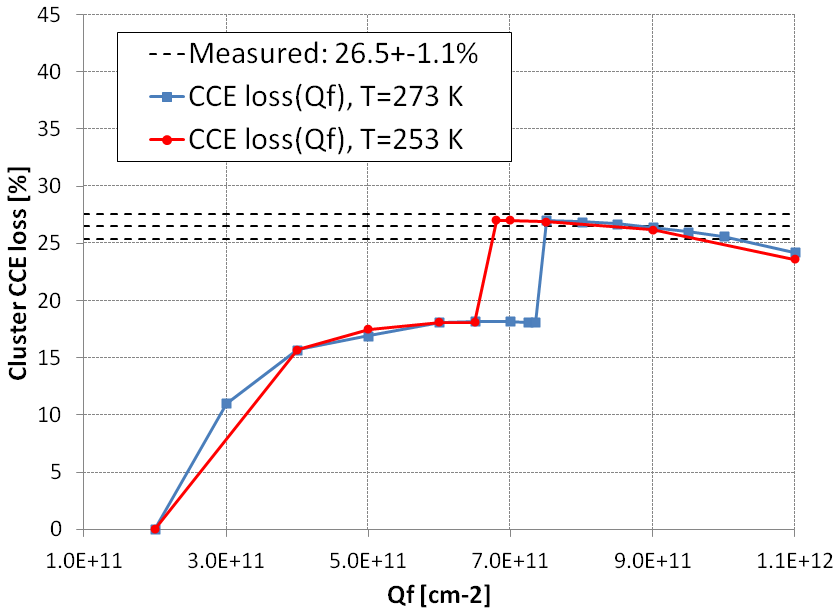}
\includegraphics[width=.47\textwidth]{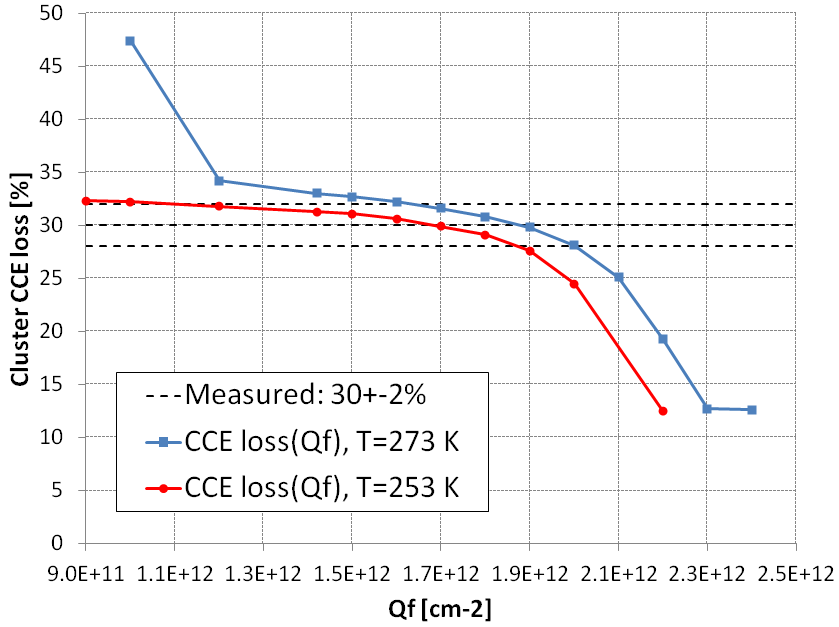}
\caption{Parametrization of the shallow acceptor concentration of the 3-level defect model as a function of fluence for 200 $\mu\textrm{m}$ active thickness n-on-p strip sensors at $T$ = 263$\pm$10 K. The strip pitch is 120 $\mu\textrm{m}$ and implant width is 28 $\mu\textrm{m}$. (left) Simulated $\Phi_{\textrm{\tiny eq}}=3\times10^{14}$ $\textrm{cm}^{-2}$ at $V$ = -1 kV. Measured detectors were proton irradiated. (right) Simulated $\Phi_{\textrm{\tiny eq}}=1.4\times10^{15}$ $\textrm{cm}^{-2}$  at $V$ = -1 kV. Measured detectors were irradiated by mixed fluences $\Phi_{\textrm{\tiny eq}}=(1.4\pm0.1)\times10^{15}$ $\textrm{cm}^{-2}$ with relative fractions n/p = $0.60\pm0.04$.}
\label{fig:6}
\end{figure}

\paragraph{} The average cluster CCE loss between the strips in 200 $\mu\textrm{m}$ active thickness float zone and magnetic-Czochralski sensors with p-spray and p-stop isolations (FZ200P/Y and MCz200P/Y, respectively), 
measured with the SiBT at $T$ = 263$\pm$10 K, was $26.5\pm1.1\%$ and $30\pm2\%$ 
\cite{bib6} as illustrated on the left and right side of figure~\ref{fig:6}, respectively. When $Q_{\textrm{\tiny f}}$ is used as an iteration parameter, the simulation produces matching cluster CCE loss at $Q_{\textrm{\tiny f}}=(8.5\pm1.0)\times10^{11}$ $\textrm{cm}^{-2}$ and $(1.6\pm0.2)\times10^{12}$ $\textrm{cm}^{-2}$, respectively, with error margins determined by propagation of errors of the curves at each simulated temperature.

Simulated CCE($x$) was found to be dependent on the shallow acceptor concentration in the 3-level defect model and on $Q_{\textrm{\tiny f}}$ at a given fluence. Thus, if one is fixed, it is possible to parametrize the other as a function of fluence. In the lack of exact measured values of $Q_{\textrm{\tiny f}}$ estimated values were used, against which the shallow acceptor concentration was tuned. Hence, at this point the approach provides more a method than quantitative information. The resulting preliminary parametrization of the shallow acceptor concentration is presented in table~\ref{tab:2}.

On the left side of figure~\ref{fig:6} the simulated cluster CCE loss in the low $Q_{\textrm{\tiny f}}$ region falls below the experimentally observed values. This is due to increased opposite sign contribution to the cluster charge from strips further away from the position of the charge injection. This is in line with the Schockley-Ramo Theory, where the negative undershoots are due to the travel distance of the charges. Thus, while drifting to the electrode, the charge induces on the other strips an undershoot. Hence, the cluster charge collected from injection at the center of the strip moves closer to the cluster charge collected from injection at the middle of the pitch.  
\begin{table}[tbp]
\caption{Three-level defect model included in the non-uniform 3-level model \cite{bib2}, parametrized for the fluence range $(0.3 - 1.5)\times10^{15}$ n$_{\textrm{\tiny eq}}\textrm{cm}^{-2}$. $E_{\textnormal{\tiny C,V}}$ are the conduction and valence band energies, $\sigma$$_{\textnormal{\tiny e,h}}$ are the electron and hole capture cross sections and $\Phi$ is the fluence.
}
\label{tab:2}
\smallskip
\centering
\begin{tabular}{lclclclclc}
    \hline
    {\small \bf{Defect}} & {\small \bf{Energy} \textnormal{[eV]}} & {\small \bf{$\sigma$$_{\textnormal{\tiny e}}$} \textnormal{[cm$^{2}$]}} & {\small \bf{$\sigma$$_{\textnormal{\tiny h}}$} \textnormal{[cm$^{2}$]}} & {\small \bf{Concentration} \textnormal{[cm$^{-3}$]}}\\
    \hline
    {\small Acceptor} & {\small $E_{\textnormal{c}}-0.525$} & {\small 10$^{-14}$} & {\small 10$^{-14}$} & {\small 1.189$\times$$\Phi$+6.454$\times$10$^{13}$}\\
    {\small Donor} & {\small $E_{\textnormal{v}}+0.48$} & {\small 10$^{-14}$} & {\small 10$^{-14}$} & {\small 5.598$\times$$\Phi$-3.959$\times$10$^{14}$}\\
    {\small Shallow acceptor} & {\small $E_{\textnormal{c}}-0.40$} & {\small 8$\times$10$^{-15}$} & {\small 2$\times$10$^{-14}$} & {\small 14.417$\times$$\Phi$+3.168$\times$10$^{16}$}\\
    \hline
\end{tabular}
\end{table}
\section{Conclusions}\label{sec:6}
\paragraph{} The two level defect models for both protons and neutrons were applied for the CCE simulations of 300 and 200 $\mu$m active thickness n-on-p strip sensors for fluences up to $\Phi_{\textrm{\tiny eq}}=1.5\times10^{15}$ $\textrm{cm}^{-2}$. By adjusting the reverse bias voltage and the interface charge it is possible to reach close agreement with CCE data measured by both ALiBaVa and SiBT set-ups. 

Implementation of realistic $Q_{\textrm{\tiny f}}$ values for proton irradiated sensors modelled by the two level model proved to be problematic since these resulted in the loss of strip isolation and high interstrip capacitances. The problem was addressed by applying the non-uniform 3-level defect model, that allowed to reach CCE($x$) matching to the data measured by the SiBT set-up. The simulated CCE($x$) was observed to be governed by $Q_{\textrm{\tiny f}}$ and the shallow acceptor concentration. By tuning these, the measured CCE loss between the strips for a given $\Phi_{\textrm{\tiny eq}}$ was reproduced. The model also made possible the use of realistic values of $Q_{\textrm{\tiny f}}$ for proton irradiation in the CCE simulations without compromising the strip isolation or the interstrip capacitance. Calibrating the simulated CCE loss between the strips with measured results led to the preliminary parametrization of the model for the fluence range $(0.3 - 1.5)\times10^{15}$ n$_{\textrm{\tiny eq}}\textrm{cm}^{-2}$.

Next stages of this work will include further parametrization of the model with larger number of data points and the investigation of the option that the non-uniform three level model could be replaced by the implementation of interface traps. Also the development of defect models up to fluences of $10^{16}$ n$_{\textrm{\tiny eq}}\textrm{cm}^{-2}$ will be required for the simulations of the pixel detectors positioned closest to the vertex at the HL-LHC.

\end{document}